# Large-Scale Arrays of Single- and Few-Layer MoS$_2$ Nanomechanical Resonators

Hao Jia[1], Rui Yang[1], Ariana E. Nguyen[2], Sahar N. Alvillar[2], Thomas Empante[2],

Ludwig Bartels[2], and Philip X.-L. Feng[1,*]

[1]*Department of Electrical Engineering and Computer Science, Case School of Engineering, Case Western Reserve University, 10900 Euclid Avenue, Cleveland, OH 44106, USA*

[2]*Department of Chemistry and Material Science & Engineering Program, University of California, Riverside, 900 University Avenue, Riverside, CA 92521, USA*

## Abstract

**We report on fabrication of large-scale arrays of suspended molybdenum disulfide (MoS$_2$) atomic layers, as two-dimensional (2D) MoS$_2$ nanomechanical resonators. We employ a water-assisted lift-off process to release chemical vapor deposited (CVD) MoS$_2$ atomic layers from a donor substrate, followed by an all-dry transfer onto microtrench arrays. The resultant large arrays of suspended single- and few-layer MoS$_2$ drumhead resonators (0.5–2μm in diameter) offer fundamental resonances ($f_0$) in the very high frequency (VHF) band (up to ~120MHz) and excellent figures-of-merit up to $f_0 \times Q \approx 3 \times 10^{10}$Hz. A stretched circular diaphragm model allows us to estimate low pre-tension levels of typically ~15mN/m in these devices. Compared to previous approaches, our transfer process features high yield and uniformity with minimal liquid and chemical exposure (only involving DI water), resulting in high-quality MoS$_2$ crystals and exceptional device performance and homogeneity; and our process is readily applicable to other 2D materials.**

---

*Corresponding Author. Email: philip.feng@case.edu.

# Introduction

Two-dimensional (2D) materials, derived from van der Waals crystals, such as graphene and atomic layers of transition metal dichalcogenides (TMDCs), have attracted tremendous attention due to their atomically-thin structures and distinct electronic, optical, and mechanical properties that lead to novel devices [1,2,3,4,5,6,7]. Suspended 2D nanostructures can play important roles, not only in characterizing physical properties (*e.g.*, elastic modulus and thermal conductivity, *etc.*) of these ultrathin crystals [8,9,10,11,12], but also in enhancing nanoelectronic device performance through decoupling from adverse substrate effects [13,14,15]. As such they enable novel nanomechanical systems [16,17,18,19,20,21,22,23,24] for promising sensor and actuator applications.

Since the successful isolation of graphene from graphite by micromechanical exfoliation [25,26], prototypical 2D nanomechanical devices have been fabricated mostly by exfoliation of small pieces of the layered bulk onto pre-patterned microtrenches [8,16,18,20,27,28,29,30,31,32]. In order to integrate electrical signal transduction (*e.g.*, for suspended field effect transistors, FETs, and resonators with electrical excitation/readout), 2D electronic devices are often released from a substrate after electron-beam lithography (EBL) followed by a sequence of metallization, wet lift-off, wet oxide (SiO$_2$) etch (*e.g.*, buffered oxide etch, BOE) and critical point drying (CPD) [13,17,23,33,34]. Alternatively, polymer-assisted transfer techniques (*e.g.*, utilizing polymethyl methacrylate, PMMA [9, 35 , 36 ], or polydimethylsiloxane, PDMS [19,21,24,37,38]) have been used to transfer exfoliated 2D flakes onto grooved substrates and across microtrenches (with/without pre-fabricated electrodes). However, the chemical processes involved in most of these methods have the potential of contaminating or degrading the 2D materials.

While micromechanical exfoliation faces challenges in fabricating large-scale arrays of suspended 2D crystals because of the limited flake size (typically tens of microns), this method has yielded some of the best quality and cleanest devices to date. Our approach, in contrast, utilizes CVD grown films that can be homogeneous across much larger area while, as we will describe below, exhibiting material properties competitive to, or even exceeding, that of the exfoliated films. Guided by the experience obtained on exfoliation samples, our work shows progress towards scalable fabrication of arrays of suspended 2D crystalline devices in a manner that incorporates: (i) improved yield by utilizing large area 2D materials, such as those obtained via CVD growth of atomic layers [39,40,41,42]; (ii) avoidance of material contamination, degradation and destruction associated with lithographic processing, chemical etching, polymer dissolution or decomposition, heating, *etc.*; (iii) compatibility with a wide set of substrate trench geometries; (iv) rapid and facile turnaround by minimal facility requirements, alignment efforts, *etc*.

Large-scale fabrication of suspended devices from CVD graphene has been reported earlier [43,44,45]. CVD graphene is usually released by wet etching from its growth substrate (*e.g.*, Cu) [43,44,45,46] followed by polymer-assisted wet- or dry-transfer [43,44,46]. These processes inevitably involve chemical etching of the growth substrates due to strong substrate interactions of the CVD films (as compared to exfoliated flakes). In addition, removal of the carrier polymer requires processing such as chemical dissolution and baking which can further contaminate and/or degrade the 2D materials of interest. In addition, solution trapping can occur, which is detrimental to device integrity and performance [47].

To address these limitations, here we describe how to quickly fabricate arrays of suspended 2D-semiconductor-on-insulator structures, by combing a water-assisted lift-off process with PDMS-based all-dry transfer. We find that by simply wetting the interface between a hydrophobic PDMS stamp and a hydrophilic SiO$_2$/Si growth substrate with droplets of deionized (DI) water, 2D thin films covering the entire sub-centimeter-scale growth substrate can be released and transferred onto the PDMS stamp (prepared on a glass slide) with no additional chemical exposure. Because of the large crystal area, no alignment is required in the subsequent all-dry transfer process from the PDMS stamp to the target substrate, further accelerating the fabrication cycle and avoiding the chance for any solution trapping. The facile nature of our process renders it feasible under almost any common lab conditions and suggests applicability to other 2D crystals that are chemically inert to DI water.

## Fabrication Process

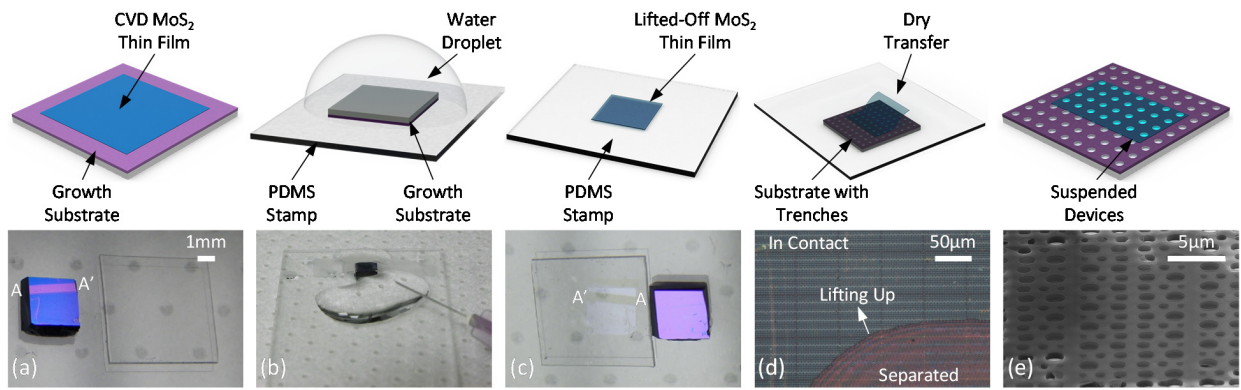

**Figure 1:** Fabrication of large-scale suspended MoS$_2$ nanomechanical structures by employing water-assisted lift-off and dry-transfer techniques. Schematic illustrations and actual images showing (a) A large-scale CVD MoS$_2$ film grown on a SiO$_2$/Si substrate (blue area is MoS$_2$ thin film, in the violet A-A' stripe the CVD MoS$_2$ has been removed). (b) Lift-off of MoS$_2$ film from growth substrate onto PDMS stamp mounted on a glass slide. DI water is injected for wetting of the SiO$_2$/MoS$_2$/PDMS interface in assistance of MoS$_2$ detachment. (c) MoS$_2$ film picked up by the PDMS stamp. The film is dried with N$_2$ gas. The color contrast and A-A' stripe in Fig. 1a and 1c indicate the successful transfer of the entire CVD MoS$_2$ film onto the PDMS stamp. (d) Dry transfer of the MoS$_2$ film onto a substrate featuring pre-patterned circular microtrench arrays. The PDMS is lifted up along the direction marked by the arrow, and the MoS$_2$ materials remain on the substrate due to stronger van der Waals adhesion to the substrate compared to the PDMS stamp. (e) A fraction of the suspended circular MoS$_2$ drumheads (of varying diameters) resulting from a single transfer cycle.

We fabricate large-scale arrays of suspended MoS$_2$ nanomechanical resonators by employing water-assisted lift-off and PDMS-based all-dry transfer. Following CVD growth of MoS$_2$ from sulfur and MoO$_3$ precursors on 290nm SiO$_2$/Si (see Methods), the growth substrate is diced into small sub-centimeter pieces with homogenous coverage of MoS$_2$. The substrate in Fig. 1a has a portion of the MoS$_2$ layer removed (violet stripe highlighted by A-A') prior to taking the image to showcase the color contrast between the bare substrate (violet stripe) and MoS$_2$-covered surface. Transfer starts by gently pressing the substrate piece onto a PDMS stamp (supported from a glass slide) and locking it in place by means of Scotch tape. Subsequently, DI water droplets are injected at the perimeter of the chip (Fig. 1b). It is critical to allow water to penetrate into the SiO$_2$/MoS$_2$/PDMS interfaces while preventing the chip from sliding during lift-off: if the growth substrate and PDMS film are pressed together so firmly that water cannot wet the interface, the MoS₂ film remains on the growth substrate. Detaching the growth substrate after 1−2min releases the MoS₂ material reliably and over a large area onto the PDMS stamp. Figure 1c shows that the whole MoS₂ film is removed and only the violet color of the virgin substrate remains; the transferred MoS₂ film can be discerned on the PDMS stamp and the A-A′ stripe is clearly visible. Subsequent transfer onto the substrate patterned with microtrench arrays (290nm in depth, 0.5−2µm in diameter) proceeds without the need for any liquid solvent (Fig. 1d). Drying of the PDMS stamp with nitrogen gas prior to the final transfer step ensures that no solution trapping (suffered by some of the previously described methods) can occur and minimizes the risk of 2D crystal degradation and damage. Figure 1e shows a fraction of the resulting array of suspended MoS₂ drumhead resonators, among the thousands obtained. The whole process is facile, rapid, and potentially applicable to a wide range of water-impervious 2D materials.

## Results and Discussions

We fabricate two types of suspended arrays by utilizing: (i) sub-centimeter size few-layer MoS₂ film and (ii) single-layer flakes (~10µm size). In each case, we have obtained extended arrays of functional nanomechanical resonators.

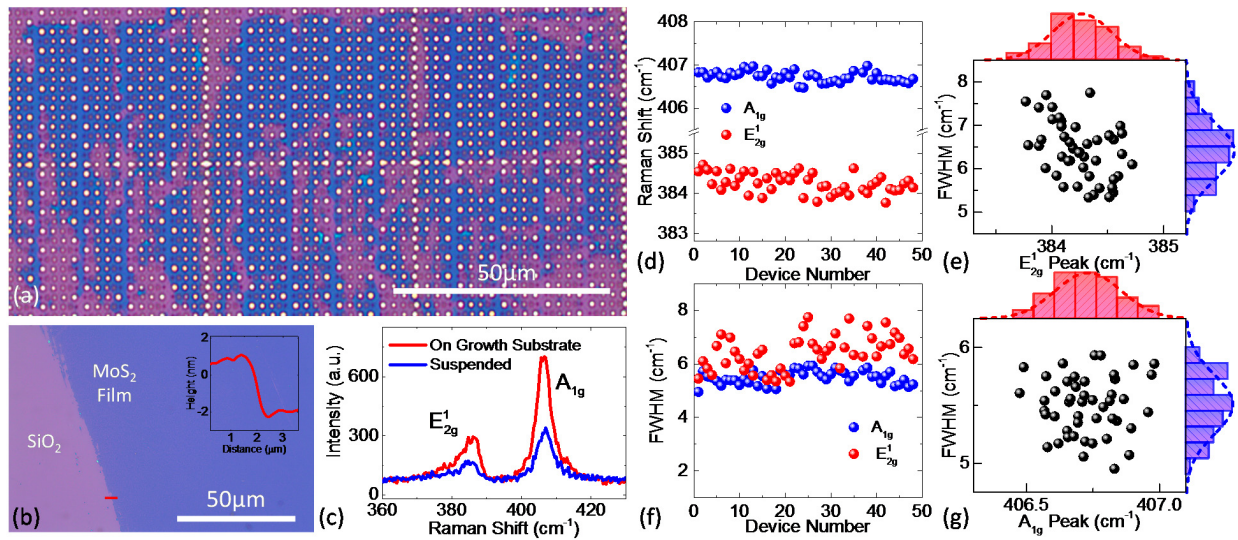

**Figure 2:** Few-layer MoS₂ nanomechanical resonator arrays. (a) Micrograph of a region containing a large number of few-layer MoS₂ nanoresonators on circular microtrenches. (b) Micrograph showing a fraction of sub-centimeter as-grown few-layer MoS₂ film. Inset: AFM step height profile showing the thickness of a 3-layer (3L) MoS₂ film (measured ~2.7nm-thick, measurement position marked by red line). (c) Comparison of Raman spectra of few-layer MoS₂ crystals on the growth substrates and suspended across microtrenches ($E_{2g}^1$ and $A_{1g}$ sharp peaks are clearly observed). (d)-(g) Statistics of Raman characteristics (peak positions and linewidths) of 47 few-layer MoS₂ drumheads ($d$=1.68µm).

Figure 2a shows an array of few-layer MoS₂ drumheads (with diameters of 0.5−2µm), which only represents a small fraction of the yield of a single transfer process. We note that the few-layer thin film (in Fig. 1c), which is orders of magnitude larger than exfoliated flakes, tore during dry transfer, forming a series of discontinuous regions of film suspended across extended arrays of substrate trenches. We attribute the tearing to mechanical deformation of the PDMS and assume that it can be avoided by further engineering of details of the transfer process. Given the large number of covered trenches (thousands) achieved in a single cycle of the current process, tearing of the MoS₂ film is no obstacle to success of this experiment.

We have examined the quality of the few-layer CVD MoS₂ film before and after fabrication using Raman spectroscopy (Fig. 2c) [40,41,48,49]: in both cases we find two sharp peaks near 380cm⁻¹ and 400cm⁻¹ corresponding to the 2 dominant phonon modes of the MoS₂ crystal, *i.e.*, $E_{2g}^1$ and $A_{1g}$. The absence of any change in linewidth of the modes in the course of the transfer process ascertains the benign nature of our procedure.

For statistical evaluation of the homogeneity of the MoS₂ few-layer film suspended across the microtrenches, we measure Raman characteristics on 47 drumheads (*d*=1.68μm) as shown in Fig. 2d-g. The $E_{2g}^1$ and $A_{1g}$ positions and linewidths are extracted by Lorentzian fitting from each spectrum (Fig. 2d and f) and their variations are fitted to normal distributions (Fig. 2e and g). We obtain $E_{2g}^1$ and $A_{1g}$ position distributions of 384.3±0.3cm⁻¹ and 406.7±0.1cm⁻¹ as well as linewidth distributions of 6.4±0.7cm⁻¹ and 5.5±0.3cm⁻¹. The separation of $E_{2g}^1$ and $A_{1g}$ peaks [40,41,49], together with the AFM height profile (inset in Fig. 2b) and optical contrast between MoS₂ and growth substrate [40,50] (in Fig. 2b) indicates a trilayer film. The standard deviations are smaller than the nominal resolution of the Raman system of ~1cm⁻¹. They confirm high uniformity of CVD MoS₂ film across the entire device regions after fabrication.

As shown in Fig. 3d, our growth substrate contains a sparse distribution of single-layer MoS₂ islands only (AFM height profile also determines single layer [19]). Our transfer technique is capable of picking up these islands quantitatively, as evidenced by the similar MoS₂ island densities on the growth substrate and the PDMS stamp shown in Fig. 3d and e. Dry-transfer onto circular microtrench arrays produces large amount of suspended single-layer MoS₂ drumheads (Fig. 3a), with single-layer flakes covering individual or multiple trenches (Fig. 3b and c). The overall yield of this process is determined by the density of single-layer islands on the growth substrate, not by the transfer process. We examine the quality of the transferred single-layer MoS₂ flakes using both Raman and photoluminescence (PL) spectroscopy [7,40,41,42,48]. The sharp $E_{2g}^1$ and $A_{1g}$ peaks in the Raman spectra (Fig. 3f) and the strong A and B exciton peaks in the PL spectra (Fig. 3g) measured before and after the transfer process indicate well-preserved crystal quality. Proceeding in the same fashion as for the few-layer films (Fig. 2), we calculate $E_{2g}^1$ and $A_{1g}$ position and linewidth distributions of 385.8±0.3cm⁻¹, 405.6±0.5cm⁻¹, as well as 2.9±0.8cm⁻¹, 4.4±0.6cm⁻¹, respectively, for 22 single-layer (*d*=0.92μm) drumheads. These measurements confirm the high and uniform quality of our single-layer devices.

Depending on the morphology of the MoS₂ films on the growth substrate, thousands of few-layer nanoresonators and hundreds of single-layer ones are prepared in a single fabrication cycle. This can still be enhanced by optimizing microtrench density and synthesizing/transferring larger, continuous single-layer films.

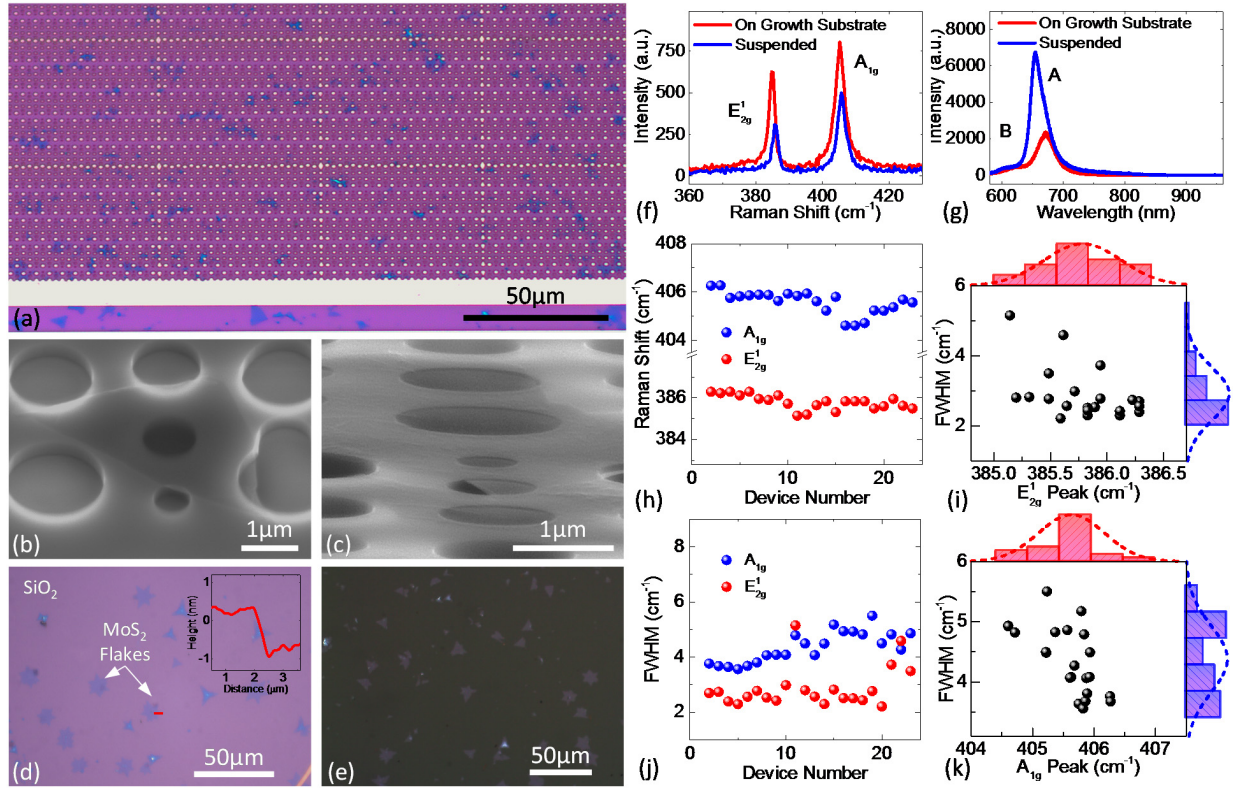

**Figure 3:** Single-layer MoS$_2$ nanomechanical resonator arrays. (a) Micrograph of region containing many MoS$_2$ nanoresonators on circular microtrenches. (b) and (c) SEM images of single-layer flakes covering single and multiple circular microtrenches, respectively. (d) and (e) Sparse distribution of single-layer CVD MoS$_2$ flakes on the growth substrate and quantitative transfer onto the PDMS stamp. Inset: AFM step height profile showing the thickness of a single-layer MoS$_2$ flake (measured ~0.9nm-thick, measurement position marked by red line). (f) Raman and (g) photoluminescence (PL) spectra of single-layer MoS$_2$ crystals on the growth substrates and suspended over microtrenches ($E^1_{2g}$ and $A_{1g}$ peaks are clearly observed in (f), and A and B peaks are observed in (g)). (h)-(k) Statistics of Raman characteristics (peak positions and linewidths) of 22 ($d$=0.92μm) single-layer MoS$_2$ drumheads.

In the following, we validate that our MoS$_2$ drumheads can readily function as resonators, and further analyze their resonant performance statistically. We characterize their fundamental-mode nanomechanical resonances using a custom-built optical interferometry system, in which an amplitude-modulated 405nm diode laser is employed to opto-thermally excite the out-of-plane vibrations of the suspended MoS$_2$ resonators, while a 633nm He-Ne laser is focused on the devices to read out their mechanical vibrations. This optical interferometry system can achieve displacement sensitivities down to fm/$\sqrt{Hz}$ levels for nanomechanical resonators constructed out of various crystalline materials [18,21].

We characterize a total of 93 few-layer devices (35, 18, 16, and 24 each for diameters of 1.68, 1.16, 0.92 and 0.64μm, respectively, as shown in Fig. 4a) and 24 single-layer devices (3, 5, 10, and 6 each for diameters of 1.68, 1.16, 0.92 and 0.64μm, respectively, as shown in Fig. 4b). All these devices are fully-covered drumheads in few- or single-layer films. The resonance spectra are fitted to a damped simple harmonic resonator model so as to extract the resonance frequency ($f_0$) and quality ($Q$) factor for each device. The $f_0$ *vs* $Q$ scatter plots are shown in the upper panels in Fig. 4a and 4b, and the statistics of $f_0$ and $Q$ are plotted in the lower panels. We observe an increase of the resonance frequency $f_0$ as the drumhead diameter $d$ decreases. For few-layer devices the values are $f_0$=61.99±14.87MHz for $d$=1.68µm, 70.64±11.11MHz for $d$=1.16µm, 86.04±32.05MHz for $d$=0.92µm, 94.31±15.53MHz for $d$=0.64µm. For single-layer devices the values are $f_0$=21.35±5.19MHz, 36.75±15.78MHz, 69.88±14.61MHz, 81.37±16.78MHz, respectively, for the same sequence of drumhead diameters. The fundamental-mode resonances of our >100 measured few- and single-layer resonators reach up to ~120MHz, *i.e.,* into the very high frequency (VHF) band. The device with the highest figure-of-merit is marked by a green symbol in each of the graphs of Fig. 4; the highest figure-of-merit $f_0 \times Q$ obtained is $3 \times 10^{10}$Hz, *on par* or better than previously reported values for individually-prepared MoS₂ drumheads from mechanical exfoliation [18, 19, 20].

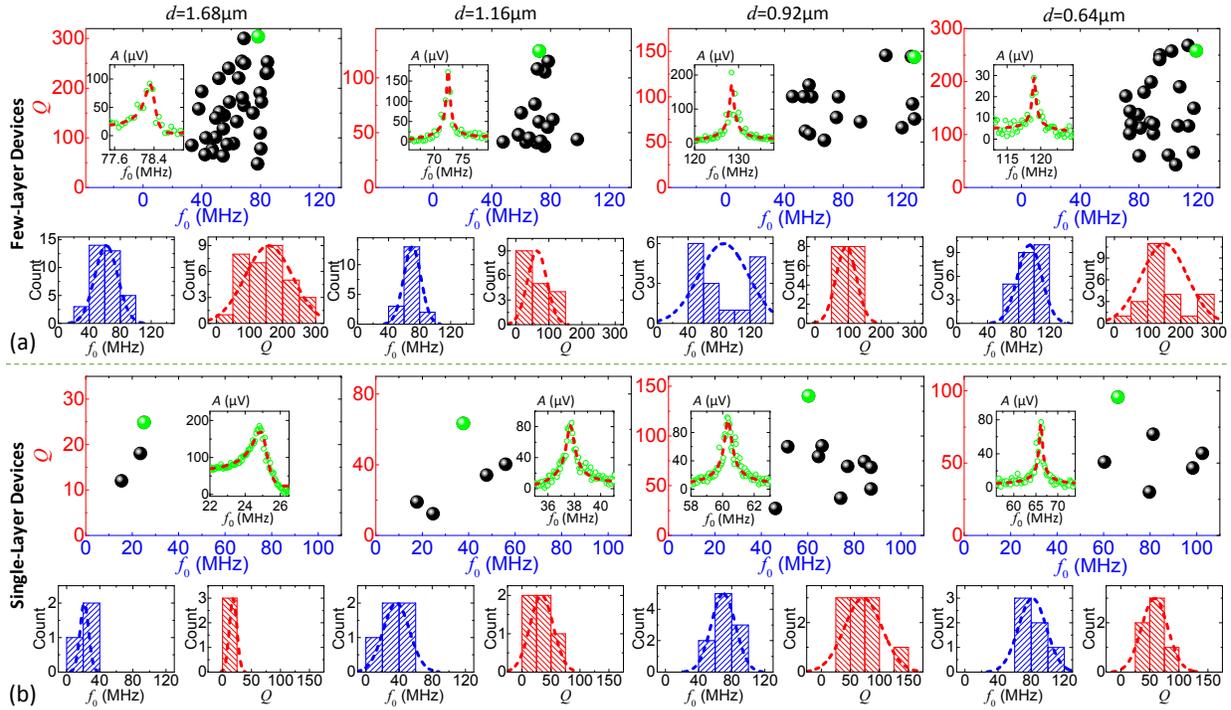

**Figure 4:** Statistics of fundamental-mode resonance characteristics. (a) and (b) depict data for few- and single-layer MoS₂ nanomechanical resonator arrays, respectively. Within (a) and (b), the upper panels show scatter plots of the fundamental-mode resonance frequency ($f_0$) *vs.* the quality ($Q$) factor for devices with 4 different diameters ($d$=1.68, 1.16, 0.92, 0.64µm). Green symbols highlight the devices with the highest figures-of-merit $f_0 \times Q$ in each graph. The insets show the resonance spectra corresponding to the green symbols as fitted to the finite-$Q$ simple harmonic resonator model (red dashed lines). The lower panels show statistics on the distributions of $f_0$ and $Q$, respectively.

We now discuss the dispersion of the fundamental-mode resonances in Fig. 4. Figure 5 compares the measured $f_0$ and $Q$ dispersions of our dataset (magenta and red) to previously reported values of exfoliated MoS₂ resonators (violet and blue) [19] and CVD graphene resonator arrays (green) [43]. We normalize the data points $X_i$ by plotting their variations from the mean $\overline{X}$ as a fraction of the mean value, *i.e.,* $(X_i - \overline{X})/\overline{X}$. Dashed lines indicate Gaussian distributions fitted to the datasets. Herein, the box symbols illustrate the dispersions, with the top and bottom of the boxes showing the dispersion percentages (*i.e.*, $\pm\sigma\langle X\rangle/\overline{X}$), and the whiskers representing the data ranges. We observe the improved $f_0$ dispersion, as low as 20.9% and 15.7% for single- and few-layer CVD MoS$_2$ resonators, in comparison to 34.1% and 22.7%, respectively, for exfoliated MoS$_2$ devices [19] and 35.8% for CVD graphene devices [43]. As for $Q$, even though we measure slightly larger dispersion in our few-layer CVD MoS$_2$ devices than in the exfoliated ones, our single-layer CVD drumheads feature an unprecedented low dispersion down to 44.3% in contrast to the 79.2% for the exfoliated MoS$_2$ devices and 57.9% for the CVD graphene devices. We attribute the improved statistical metrics of our devices to the facile and clean transfer process that maintains high and uniform crystal quality.

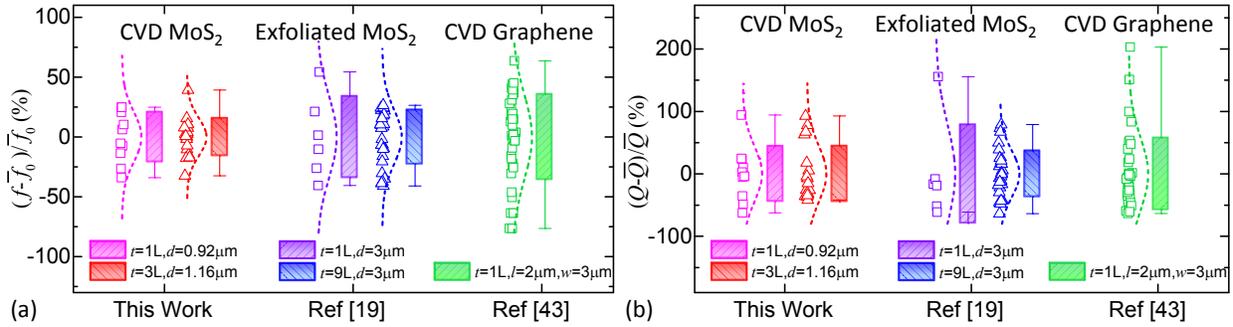

**Figure 5:** Comparison of (a) fundamental-mode resonance frequency ($f_0$) and (b) quality ($Q$) factor dispersions with published work [19,43]. The CVD MoS$_2$ nanoresonator arrays in this work are shown in magenta for single-layer and red for few-layer devices; previously reported exfoliated MoS$_2$ resonators are shown in violet for single-layer and in blue for multilayer [19]; large-scale CVD graphene resonator arrays are shown in green [43]. Both our data and the reference data are normalized as described in the text.

We also note that because nanomechanical resonances are much more sensitive (than, for instance, Raman spectra) to variations in the stiffness and effective mass of the suspended devices (*e.g.*, very small variations in the pre-tension levels that are invisible in Raman measurement can affect the resonance frequency $f_0$ quite pronouncedly), there are noticeable dispersions in $f_0$ and $Q$ values (in Fig. 4), while there is no appreciable non-uniform Raman peak position and FWHM (in Fig. 2 and 3).

Information on the pre-tension levels of our 2D MoS$_2$ drumheads can be obtained by fitting the data to a resonance frequency model of a fully-clamped stretched circular diaphragm [51]:

$$f_0 = \omega_0/(2\pi) = \left(\frac{kd}{4\pi}\right)\sqrt{\frac{16D}{\rho d^4}\left[\left(\frac{kd}{2}\right)^2 + \frac{\gamma d^2}{4D}\right]},\tag{1}$$

where $\rho$ is the areal mass density of MoS$_2$, $\gamma$ the pre-tension level (N/m), $D$ the bending rigidity $D = E_Y t^3 / \left[12\left(1-\nu^2\right)\right]$, with $E_Y$, $\nu$ and $t$ being the Young's modulus, Poisson's ratio and thickness of MoS$_2$. Here the mode parameter is determined by $\left(kd/2\right)^2 = \alpha + \left(\beta - \alpha\right)e^{\left[-\eta e^{\delta\ln(x)}\right]}$, with $x = \frac{\gamma d^2}{4D}$, $\alpha = 5.7832$, $\beta = 10.215$, $\eta = 0.1148$ and $\delta = 0.4868$, respectively [51]. Figure 6a shows the behavior of Eq. (1) for different pre-tension levels. This model explicitly incorporates contributions from both flexural rigidity (the first term, with $D$) and pre-tension (the second term, with $\gamma$) [18]. It asymptotically simplifies into a 'disk' or a 'membrane' model, when the $D$ term or the $\gamma$ term dominates, respectively. When neither term is dominating over the other, the device is in the transition regime (as quantitatively depicted in Fig. 6b).

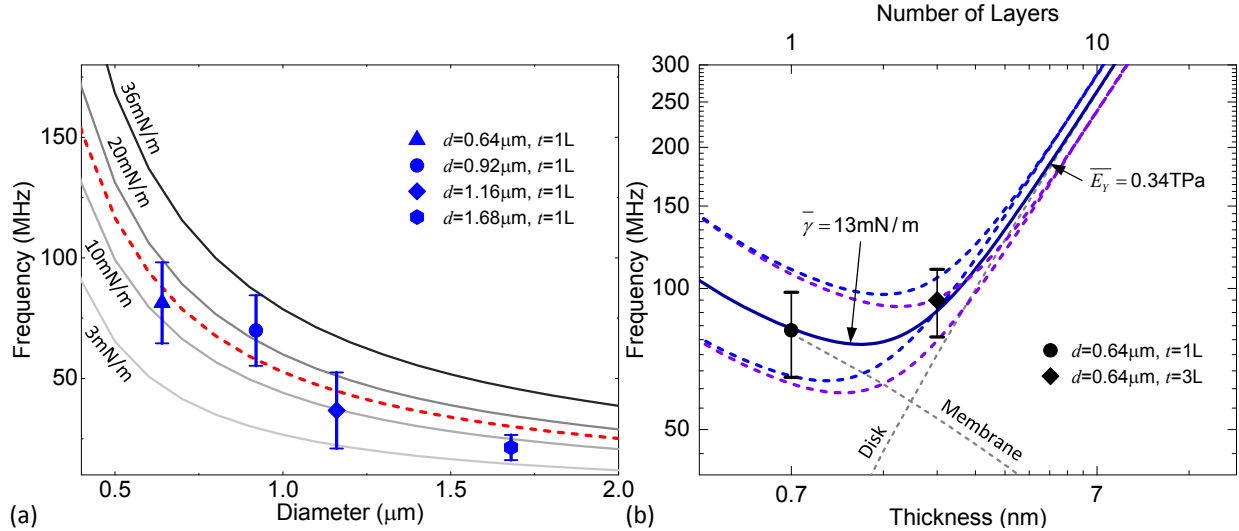

**Figure 6:** Analysis of pre-tension level and the elastic modulus of few- and single-layer MoS₂ drumhead arrays. (a) Grey curves illustrates the dependence of $f_0$ on the device diameter ($d$) under different pre-tension levels for single-layer devices by employing a mixed-elasticity model for stretched circular diaphragms. Blue symbols and whiskers highlight the mean ($\overline{f_0}$) and standard deviation ($\sigma \langle f_0 \rangle$) of the fundamental-mode resonances of single-layer device families in Fig. 4. The red dashed line indicates a best overall fit with $\gamma \approx 15$mN/m for the single-layer devices. (b) Elastic transition between the 'disk' and 'membrane' limits (gray dashed lines). The single-layer devices of $d \sim 0.5-2\mu$m (*e.g.*, $d=0.64\mu$m in (b)) are in the 'membrane' regime, while few-layer devices with the same diameters are more 'disk'-like.

Fitting the experimental results of the $f_0$ spread of single-layer drumheads in Fig. 4 (symbols and whiskers highlighting the mean and standard deviation) to Eq. (1), we obtain a best overall fit for $\gamma \approx 15$mN/m; the entire spread of the $f_0$ distribution corresponds to a pre-tension range of $\gamma \approx 3-36$mN/m (Fig. 6a). Assuming that our transfer method introduces similar pre-tension levels $\gamma$ to both single- and few-layer drumheads of the same diameter, we estimate $\gamma$ and $E_Y$ variances. Figure 6b shows the results for the $d=0.64\mu$m device family. The fitting leads to an averaged pre-tension level $\overline{\gamma} \approx 13$mN/m and Young's modulus $\overline{E_Y} \approx 0.34$TPa (solid dark blue line in Fig. 6b). The $f_0$ dispersions are consistent with $\gamma$ in a range of 7–25mN/m and $E_Y$ in a range of 0.28–0.40TPa (violet and blue dashed lines) in our samples. The extracted $E_Y$ values are in good agreement with those in previous work [47,52]. Interestingly, from the modeling we obtain much lower pre-tension levels than those reported in suspended exfoliated/CVD MoS₂/graphene structures [8,9,10,52, 53,54]. Lower pre-tension levels may be desirable in certain situations where excellent homogeneity and wide tuning ranges are required. This may be interpreted as another advantage of the water-assisted release and the dry transfer method employed in the present work.

## Conclusion

We have demonstrated fabrication of large arrays of suspended $MoS_2$ nanomechanical resonators by means of a combination of water-assisted lift-off and all-dry transfer. Our clean and efficient process releases large areas of 2D crystals from their growth substrate while only requiring DI water. This avoids the need for chemical treatment of the 2D materials and mitigates their degradation. Results from more than 100 single- and few-layer $MoS_2$ drumhead nanoresonators show fundamental resonances in the VHF band with figure-of-merit as high as $f_0 \times Q \approx 3 \times 10^{10}$ Hz. We observe higher uniformity of the frequency and quality factor, as well as lower pre-tension levels, compared to previous results. The facile nature and the absence of any solvents but DI water in our transfer technique holds promises for enabling suspended 2D devices and arrays using other 2D materials and their heterostructures, and it is applicable to substrates with other geometries of microtrenches and electrodes.

**Acknowledgment:** We thank the support from Case School of Engineering, National Academy of Engineering (NAE) Grainger Foundation Frontier of Engineering (FOE) Award (FOE2013-005), National Science Foundation (NSF) CAREER Award (Grant ECCS-1454570), and CWRU Provost's ACES+ Advance Opportunity Award. $MoS_2$ film fabrication was supported by C-SPIN, part of STARnet, a Semiconductor Research Corporation program sponsored by MARCO and DARPA, and NSF under grant ECCS-1435703. Part of the device fabrication was performed at the Cornell NanoScale Science and Technology Facility (CNF), a member of the National Nanotechnology Infrastructure Network (NNIN), supported by the National Science Foundation (Grant ECCS-0335765). A.E.N. gratefully acknowledges fellowship support under NSF DGE-1326120. H.J. acknowledges fellowship support under China Scholarship Council (No. 201306250042). H.J. and P.X.-L.F. thank Jaesung Lee for helpful discussions.

# Methods

## CVD MoS₂ Synthesis

The $MoS_2$ material was synthesized via CVD utilizing elemental sulfur and $MoO_3$ powder as precursors. The process is described in Ref. [55] and the resultant material compares well with exfoliated material [56]. In brief, film growth proceeds by placing alumina boats containing the precursors at different positions in a quartz process tube. The growth substrate is placed directly atop the boat containing the $MoO_3$ powder. Initially, the quartz tube is inserted into a tube furnace so that the $MoO_3$ boat comes to rest at the furnace's center. The furnace is gradually heated up to 680°C while a continuous flow of $N_2$ gas transports the sulfur vapor to the growth substrate. After the furnace has reached the growth temperature, the temperature is held for a few minutes and then the furnace is cooled gradually to room temperature while the $N_2$ gas flow is maintained. This procedure results in a few-layer $MoS_2$ film where the growth substrate covers the $MoO_3$ boat; where the growth substrate overhangs the $MoO_3$ boat on the side, single-layer film and isolated single-layer flake islands are obtained. The $MoS_2$ crystals were characterized optically via Raman and Photoluminescence (PL) spectroscopy before being shipped from UC Riverside to Case Western Reserve University for further Raman and PL measurements, transfer, and the following characterization of the drumheads and arrays.

## Fabrication of Microtrenches

Large arrays of microtrenches are lithographically defined and fabricated on 290nm-$SiO_2$-on-Si substrate. The $SiO_2$ layer is patterned with microtrenches using photolithography, and then etched by reactive ion etching (RIE). The depth of resulting $SiO_2$ microtrenches is 290nm in this experiment, and the nominal values for the diameters in the design are 1.5, 1, 0.75 and 0.5μm (corresponding to 1.68, 1.16, 0.92, 0.64μm after the fabrication process).

## Interferometric Resonance Measurement

Fundamental resonances of the resulted suspended $MoS_2$ drumhead resonators are characterized using optical excitation and detection techniques. An amplitude-modulated 405nm diode laser is employed to opto-thermally excite the flexural vibrations of the suspended $MoS_2$ nanoresonator, while a 633nm He-Ne laser is focused on the device surface for detection of the motion-modulated interference between multiple reflections from the flake-vacuum, vacuum-$SiO_2$ and $SiO_2$-Si interfaces, which is read out by a low-noise photodetector. A network analyzer is used to modulate the blue laser as well as measure the response from the photodetector. We apply a blue laser power of ~39μW and red laser power of ~1.23mW to the devices which are preserved under moderate vacuum conditions (~90mTorr) to ensure reasonable signal-to-noise ratio while avoiding laser degradation and overheating.

## Scanning Electron Microscopy (SEM), Atomic Force Microscopy (AFM), and Raman/Photoluminescence (PL) Spectroscopy

SEM images are taken inside an FEI Nova NanoLab 200 field-emission SEM, using an acceleration voltage of 10kV. AFM measurements are conducted with a Park NX10 AFM using tapping mode. Raman and PL spectra are taken by employing a Horiba iHR550 system using gratings with groove densities of 2400g/mm and 1200g/mm and integration times of 30s and 10s, respectively.